\documentclass[
reprint,
superscriptaddress,
nofootinbib,
 amsmath,
 amssymb,
aps,
]{revtex4-2}

\usepackage{graphicx}
\usepackage{dcolumn}
\usepackage{bm}
\usepackage{hyperref}
\usepackage{enumitem}
\usepackage{amsmath}

\begin{document}
\preprint{APS/123-QED}

\title{Complex multiple-choice questions are inequitable for low-income and domestic students (and difficult for everyone)}

\author{Nicholas T. Young}
\email{Corresponding author: ntyoung@umich.edu}
\affiliation{Center for Academic Innovation, University of Michigan, Ann Arbor, MI, 48104, USA}
\author{Mark Mills}
\affiliation{Center for Academic Innovation, University of Michigan, Ann Arbor, MI, 48104, USA}
\author{Rebecca L. Matz}
\affiliation{Center for Academic Innovation, University of Michigan, Ann Arbor, MI, 48104, USA}
\author{Eric F. Bell}
\affiliation{Department of Astronomy, University of Michigan, Ann Arbor, MI, 48105, USA}
\author{Caitlin Hayward}
\affiliation{Center for Academic Innovation, University of Michigan, Ann Arbor, MI, 48104, USA}

\date{\today}

\begin{abstract}
High-stakes exams play a large role in determining an introductory physics student’s final grade. These exams have been shown to be inequitable, often to the detriment of students identifying with groups historically marginalized in physics. Given that exams are made up of individual questions, it is plausible that certain types of questions may be contributing to the observed equity gaps. In this paper, we examine whether that is the case for one type of forced-choice question, the complex multiple-choice (CMC) question. In a CMC question, students must select an answer choice that includes all correct responses and no incorrect responses from a list. To conduct our study, we used four years of data from Problem Roulette, an online program at our university that allows students to prepare for exams with actual questions from previous years’ exams in a not-for-credit format. We categorized the 951 Physics II (Electricity and Magnetism) questions in the database as CMC or non-CMC. We found that students performed 8.0 percentage points worse on CMC questions than they did on non-CMC questions. In addition, we found differential impacts for low income students and domestic students relative to medium and high income students and international students. Regression models supported these descriptive findings. The results suggest that complex multiple-choice questions may be contributing to the equity gaps observed on physics exams. Considering, however, that CMC questions are more difficult for everyone, future research should examine the source of this difficulty and whether that source is functionally related to learning and assessment. For example, our data does not support using CMC questions instead of non-CMC as a way to differentiate top-performing students from everyone else.
\end{abstract}

\maketitle

\section{Introduction}\label{sec:intro}
Despite historical and ongoing efforts, many science, technology, engineering, and mathematics (STEM) fields have yet to achieve gender and racial parity \cite{national_center_for_science_and_engineering_statistics_ncses_diversity_2023}. Physics has been one of the least successful STEM disciplines in increasing its diversity, with the percent of women earning physics degrees stagnating around 20\% \cite{american_physical_society_bachelors_2021} and the percent of degrees awarded to African American, Hispanic American, American Indian, Alaska Native, Native Hawaiian, and other Pacific Islander students
\footnote{Throughout this paper when referencing another publication, we will use the terminology used by that publication for referring to a student group. Therefore, terms that may appear to be used interchangeably should not necessarily be interpreted as such without consulting the source publication.}
even lower \cite{american_physical_society_degrees_2021}. As producing a sufficient number of STEM graduates is a national priority for the United States \cite{chen_stem_2013, holdren_engage_2012}, efforts must continue to focus on attracting and retaining students from groups marginalized in STEM.

One of the most studied areas regarding student retention is the impact of introductory courses given their outsized impact \cite{seymour_talking_2019}. Grades may be interpreted by students to reflect what they are good at and can inform whether they should remain in or change their major \cite{ost_role_2010, rask_attrition_2010, stinebrickner_major_2014}. Yet, grades are not assigned equally across student groups. For example, a multi-institutional study found that men tended to earn higher grades than women did in introductory physics relative to their other courses \cite{matz_patterns_2017}, another study found that men tended to earn higher grades than women did across the physics curriculum \cite{malespina_gender_2022}, and a third study found that students of color, first-generation students, and low-income students tended to earn lower grades than middle- and upper-class, continuing-generation, white, and Asian students \cite{whitcomb_not_2021}. This last study further found that even the most advantaged students of color earned lower grades than the most disadvantaged white and Asian students.

These grade differences impact who earns a STEM degree as prior work has found that earning low grades early on in STEM courses decreases the chance that a student will graduate with a major in STEM \cite{chen_stem_2013, seymour_talking_2019, thompson_grade_2021}. In addition, the effect is not the same for all students as studies have found that women who failed a key ``weed-out" STEM course were less likely to graduate with a STEM degree than men who failed the same course \cite{sanabria_weeded_2017}. More broadly, men appear to be less sensitive to feedback about their actual course performance and hence, remain in STEM fields despite having lower grades than women who left STEM \cite{cimpian_understanding_2020}. Other work has found that the relationship between earning a lower grade in introductory STEM courses and not earning a STEM degree was stronger for students from minoritized in comparison to majority races and ethnicities \cite{hatfield_introductory_2022}.

Yet, not all courses are equal when it comes to grading disparities. Even if lecture courses in biology, chemistry, and physics perpetuate grade inequities, prior work shows that the corresponding laboratory courses do not reflect the same grade inequities \cite{matz_patterns_2017}. The authors of that study reasoned that that result was likely due to the assessment mechanics within lecture and lab courses, where lecture courses tend to use high-stakes, timed, multiple-choice exams while labs tend to use written reports, projects, and low-stakes quizzes. 

In introductory physics specifically, high-stakes exams often make up a substantial portion of the final grade \cite{koester_patterns_2016, simmons_grades_2020} and are generally made up of forced-choice questions. Performance on these exams is inequitable. Prior work has found that men tend to outperform women on physics exams \cite{kost_characterizing_2009, kost-smith_gender_2010} and more generally in STEM courses \cite{salehi_gender_2019, cotner_can_2017}. 

\begin{figure}
    \centering
    \includegraphics[width=.7\linewidth]{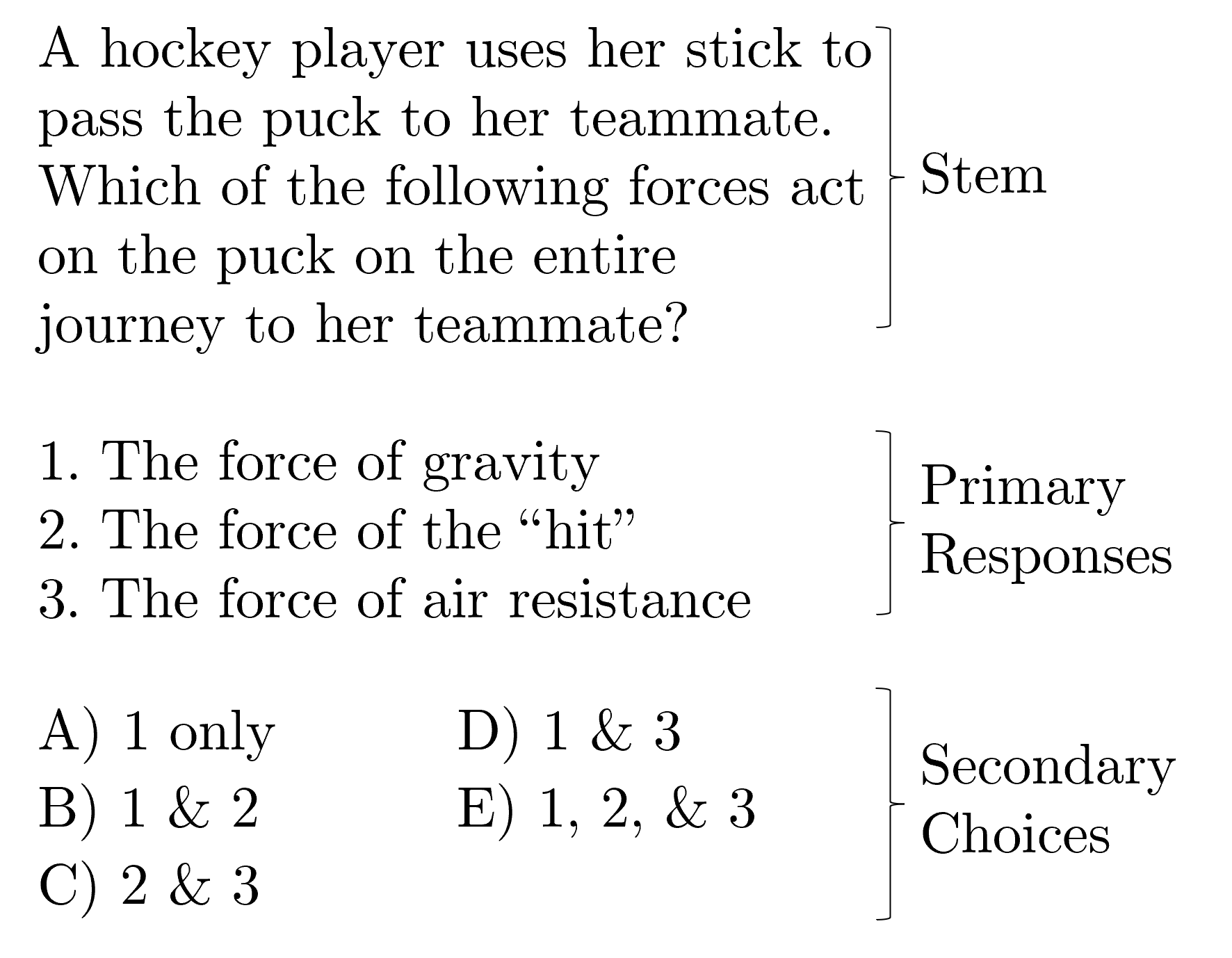}
    \caption{A sample complex multiple-choice question indicating the stem, primary responses, and second choices.}
    \label{fig:sample_CMC}
\end{figure}

Given that exam grades have been found to be inequitable in physics, it is possible that individual multiple-choice questions on exams are also inequitable and in sum, produce some part of the overall grade inequities observed on exams. For example, various physics studies have noted that the presentation and context of a question, independent of the underlying physics content, can lead to gendered performance differences \cite{mccullough_differences_2001,mccullough_gender_2011, low_persistent_2015}. In this paper, we examine a specific type of multiple-choice question, the complex multiple-choice (CMC) question in which the student is given multiple possible responses to a question and must select the answer choice including all correct responses (see Fig. \ref{fig:sample_CMC} for an example). CMCs are assumed to require and measure higher-order thinking skills \cite{berk_consumers_1996}, but at the same time may add unnecessary cognitive load unrelated to the physics content and potentially benefit some students for reasons unrelated to their physics knowledge.

In this paper, we ask if CMC questions exacerbate inequity in introductory physics courses or equivalently, if complex multiple-choice questions disproportionately benefit some students compared to others. We break this overarching question into five sub-questions:
\begin{enumerate}
    \item How do students' performance on CMC questions compare to their performance on non-CMC questions in introductory physics?
\end{enumerate}

Because the format of CMC questions allows for multiple correct answers, CMC questions often test concepts rather than calculations. We assume this to be the case for all CMC questions in this study. As prior work has found that students often perform better on ``plug-and-chug" calculations questions compared to conceptual questions \cite{kohl_student_2005}, it is possible that any performance gaps found could be a result of that. Therefore, to clarify the answer to question one, we additionally ask:

\begin{enumerate}[resume]
    \item How does the type of knowledge students need to answer a question (conceptual vs. numerical) account for performance differences between CMC questions and non-CMC questions?
\end{enumerate}

Finally, to determine how CMC questions may exacerbate inequity in introductory physics courses, we ask three additional questions: 

\begin{enumerate}[resume]
    \item What is the relationship between students' prior preparation and their ability to answer CMC questions correctly?
    \item How well can a CMC question differentiate between students who go on to earn a high versus low grade in their physics course?
    \item How might CMC questions affect performance gaps between students of various demographic groups?
\end{enumerate}

The rest of the paper is organized as follows. In Sec. \ref{sec:background}, we provide an overview of existing research on complex multiple-choice questions, both in general and in physics specifically. In Sec. \ref{sec:methods}, we describe how we obtained the data from our Problem Roulette database and how we created the groups used in our analysis. In Sec. \ref{sec:results}, we show the results of descriptive statistics and regression models finding that CMC questions are harder for students. In Sec. \ref{sec:RQs}, we compare the results obtained by the two analytical methods. In Sections \ref{sec:limits} and \ref{sec:future}, we consider how our data source limits the generalizability of the results and suggest directions for future work.

\section{Background}\label{sec:background}
\subsection{Complex multiple-choice questions}
The complex multiple-choice (CMC) question format was originally developed by the Educational Testing Service and used extensively by the National Board of Medical Examiners for medical student examinations \cite{haladyna_effectiveness_1992}. When used on these exams, CMC questions often appear in the "type-K" or "K-type" format, which is a subcategory of CMC questions with four primary responses and five of the possible sixteen responses displayed as secondary choices \cite{albanese_type_1993}. As a result of appearing on these types of assessments, many CMC studies have occurred in the context of medical education and standardized tests. To be considered a CMC, the question must have a stem, a set of primary responses, and a set of secondary choices that contain various combinations of the primary responses \cite{albanese_type_1993}. Multiple-choice questions that do not follow this format, even if they require a student to select multiple answer choices, are not considered complex multiple-choice. In this paper, we do not differentiate between CMC questions and type-K questions.

Complex multiple-choice questions are often used instead of more traditional multiple-choice questions due to the assumed benefits they provide. For example, the higher complexity of the question is assumed to enable assessment of greater complexity in thinking \cite{butler_multiple-choice_2018}. Given that traditional multiple-choice questions have been claimed to hinder critical thinking skills in introductory science courses \cite{stanger-hall_multiple-choice_2012}, complex multiple-choice questions could provide a remedy while retaining the benefits of multiple-choice questions such as being easy to grade and implement in large-enrollment courses. 

One specific benefit may be the ability to identify when students believe more than one response is true. For example, a biology education study found that as many as half of students who correctly answered a traditional multiple-choice question would have also (incorrectly) selected another answer choice if given the option \cite{couch_multipletruefalse_2018}. While that study was done with multiple-true-false questions, where a student answers true or false individually to each primary response rather than select the combination of true statements, instead of complex multiple-choice questions, it suggests that traditional multiple-choice questions might not be able to capture the partial understandings student might hold. The CMC format can also be useful for assessing knowledge in cases when there is more than one correct answer. In the context of physics, that could include identifying all forces acting on an object or answering lab safety questions.

Despite the proposed benefits of this type of question, complex multiple-choice questions also have been documented to have disadvantages relative to traditional multiple-choice questions. First, multiple studies in disciplines such as educational assessment, mathematics, and medicine have found that CMCs are harder for students than traditional multiple-choice questions \cite{haladyna_validity_1989, nnodim_multiple-choice_1992, albanese_type_1993, ozkan_student_2018}, though the effect may depend on whether the question requires the recall of memorized information or higher-order thinking skills \cite{tripp_are_1985}. As a result, some have claimed that the reason for the difference in performance could be because CMCs are measuring more than just knowledge such as non-cognitive traits \cite{nnodim_multiple-choice_1992}. For example, after reformatting complex multiple-choice questions as multiple-true-false questions, one study found that student performance improved \cite{nnodim_multiple-choice_1992}.

Second, CMCs may inadvertently ``clue" students to the correct answer \cite{albanese_type_1993} because not all combinations of primary responses can be present in the secondary choices if there are five or fewer answer choices (a common practice on exams). For example, in Figure \ref{fig:sample_CMC}, knowing that response 2 is incorrect eliminates answers B, C, and E. Of the remaining options, both have response 1. Therefore, a student is able to get the question correct only by knowing whether responses 2 and 3 are true or false.  This clueing could then help less knowledgeable students perform better on complex multiple-choice questions than they would on comparable traditional multiple-choice questions \cite{albanese_type_1993}.

Third, complex multiple-choice questions might be inequitable, though the evidence is mixed. One study of undergraduates in a teaching certification program found that some, but not all complex multiple-choice questions in their sample showed performance gaps between male and female students \cite{kaplan_examining_2019} while a different study with students enrolled in either an accounting or law course found no such differences \cite{delaere_performance_2011}.

Finally, there are other concerns about CMC questions from a test construction perspective. CMC questions have been shown to have lower reliability than traditional multiple-choice questions or multiple-true-false questions \cite{berk_consumers_1996}. In addition, CMC questions can be harder to construct \cite{berk_consumers_1996}, though others disagree and believe that they require fewer distractor answers than traditional questions \cite{manoharan_cheat-resistant_2019}. For example, a test creator could make five secondary choices from a CMC question with only one correct response and two incorrect responses, while a corresponding traditional question would require four incorrect responses.

Given the number of concerns about CMC questions, many have recommended against using the format \cite{albanese_type_1993, haladyna_review_2002, butler_multiple-choice_2018}. Yet, recent work suggests that this format is still in use \cite{rustanto_developing_2023, mcfarland_development_2017, siswaningsih_implementation_2023, stowe_you_2021}.

\subsection{Complex multiple-choice questions in physics}
To our knowledge, there has been limited study of the CMC question format in the physics context. One paper \cite{malik_effect_2020} found that students performed better on CMC questions than on equivalent versions in the multiple-true-false format with around one-third of students answering the multiple-true-false questions in a way that was not represented among the secondary choices in the corresponding CMC format. The study concluded that the CMC format may overestimate student knowledge.

Despite the limited direct study of CMC questions in physics, there have been many indirect studies due to their appearance on many concept inventories including the Test of Understanding of Graphs in Kinematics (items 12 and 19) \cite{beichner_testing_1994}, the Force Concept Inventory (items 12 and 22 on the original 1992 version; items 5, 18, 29, and 30 on the 1995 revision) \cite{hestenes_force_1992}, and the Physics Inventory of Quantitative Literacy (item 4) \cite{white_brahmia_physics_2021}. In the case of the 1995 version of the Force Concept Inventory, various studies have commented on the CMC items, noting that items 5 and 18 were able to distinguish between high- and low-proficiency students \cite{wang_analyzing_2010}; items 5, 18, and 29 are ``problematic" \cite{traxler_gender_2018}; and item 29 is ``potentially malfunctioning for currently unknown reasons" \cite{eaton_generating_2019}. Thinking toward the next generation of concept inventories in physics (e.g., Laverty and Caballero \cite{laverty_analysis_2018}), it is important to consider whether CMC-type questions should be avoided on these types of assessments.

CMC questions in physics have also appeared in studies of in-class interventions \cite{hu_challenges_2022}, of K-12 contexts \cite{kumas_assessment_2021}, and of teacher knowledge assessment \cite{vogelsang_development_2020}. Outside of peer-reviewed studies and conceptual inventories, CMC questions have also appeared on the American Association of Physics Teachers' $F=ma$ exam, which is used as the first qualifier for the US Physics Team in the International Physics Olympiad and on the physics Graduate Record Examination (GRE) which, until recently, was required for most applicants to physics graduate programs \cite{guillochon_gre_nodate}. These include the released 2001 and 2008 exams and the Educational Test Service's GRE Physics Test Practice Book \cite{educational_testing_service_gre_2020}.

\section{Methods}\label{sec:methods}
\subsection{Data Collection}\label{sec:data}
Data for this study comes from Problem Roulette users enrolled in a second-semester calculus-based introductory physics course (referred to as Physics II herein) between the Winter 2019 and Winter 2023 semesters (13 semesters total, including summer semesters which enroll significantly fewer students than Fall and Winter offerings). Problem Roulette is a free, optional, not-for-credit, test question practice software offered to all students in introductory physics courses (and other introductory STEM courses) \cite{evrard_problem_2015}. Students have the option to select a broad physics topic or specific exam period (e.g., Exam 1) and then are randomly served practice questions from the system. Students are given as many attempts as they need to answer the question and are told whether they are correct or not after each response. All questions are uploaded by instructors and appeared on exams during previous course iterations. Prior work has found that using this system regularly results in a significant increase in a student's final grade in introductory physics over and above what might be expected based on their prior academic preparation (i.e., grades in their other college courses and ACT/SAT scores) \cite{evrard_problem_2015, black_quantifying_2023}. For the time period in this study, 1685 students answered at least one question in Problem Roulette, corresponding to 41\% of the students enrolled in the course and 39\% of students who completed the course during the study period.

Data for this study also came from a simplified version of the university's student data warehouse intended to provide researchers with straightforward access to student data \cite{lonn_rearchitecting_2019}. Specifically, we had access to students' course grades, birth sex, race/ethnicity, high school zip code, parental education status, and residency status.

\subsection{Demographic definitions}\label{sec:demo}
Noting that demographic categories are neither natural nor given \cite{gillborn_quantcrit_2018}, in this section, we discuss how we defined each demographic group. As the demographic data is collected by the university and not directly by us as researchers, we had no control over the options students were able to select from and acknowledge that they are not the only axes on which privilege and oppression act in physics. The five demographic categories were sex, race/ethnicity, socioeconomic status, first-generation status, and transfer status. 

Sex was recorded as a binary variable (male or female) even though sex is not binary \cite{westbrook_new_2015} and gender is actually the construct of greater interest. 

For race/ethnicity, students were able to select any of the following five options: Asian, Black, Hispanic, Native American, and white. Any student who selected more than one was marked as Multiracial. Due to the limited numbers of Black, Hispanic, Multiracial, and Native American students in our study population, we grouped these students into one category and we grouped Asian and white students into a second category. Given that the underrepresented minority label is considered racist and harmful \cite{bensimon_misbegotten_2016, walden_critiquing_2018, williams_underrepresented_2020}, we instead refer to these groups as ``B/H/M/N" and ``A/w", using the first letter of each race and ethnicity recorded by the university, following the capitalization used by Robertson et al. in a recent paper in this journal \cite{robertson_race-evasive_2023}. We acknowledge that such groupings ignore the unique situations of each group and further mask the struggles of individuals \cite{teranishi_race_2007, bensimon_misbegotten_2016, shafer_impact_2021}. 

We defined a student as low-income with a binary variable where a student was considered low-income if the median annual household income of the student's high school's zip code was less than \$60,000. We determined median annual household income by zip code using the 2020 American Community Survey 5-Year Estimates.  We used this metric rather than the student's self-reported estimated family income---another socioeconomic metric available to us---due to a large amount (more than 20\%) of missing data and a lack of data around which students received Pell Grants, a common choice for denoting low-income students in education research. We also use this community-based definition of low income rather than an individually-based definition to acknowledge that educational resources students have access to are not solely based on their family but also their community. Therefore, a low-income student based on family income may have had access to different opportunities if they attend a well-resourced school compared to a low-income student based on family income who attends a less-resourced school. Students without a zip code listed (such as international students) were conservatively coded as not low-income. 

We defined first-generation status as a binary variable where any student whose parents did not earn at least a bachelor's degree was marked as a first-generation student. 

International students' status was based on whether the university considers the student to be international, defined as students who have a citizenship status of Non-Resident Alien or Alien Under Tax Treaty.

\subsection{Analysis}\label{sec:analysis}
\begin{figure}
    \centering
    \includegraphics[width=.9\linewidth]{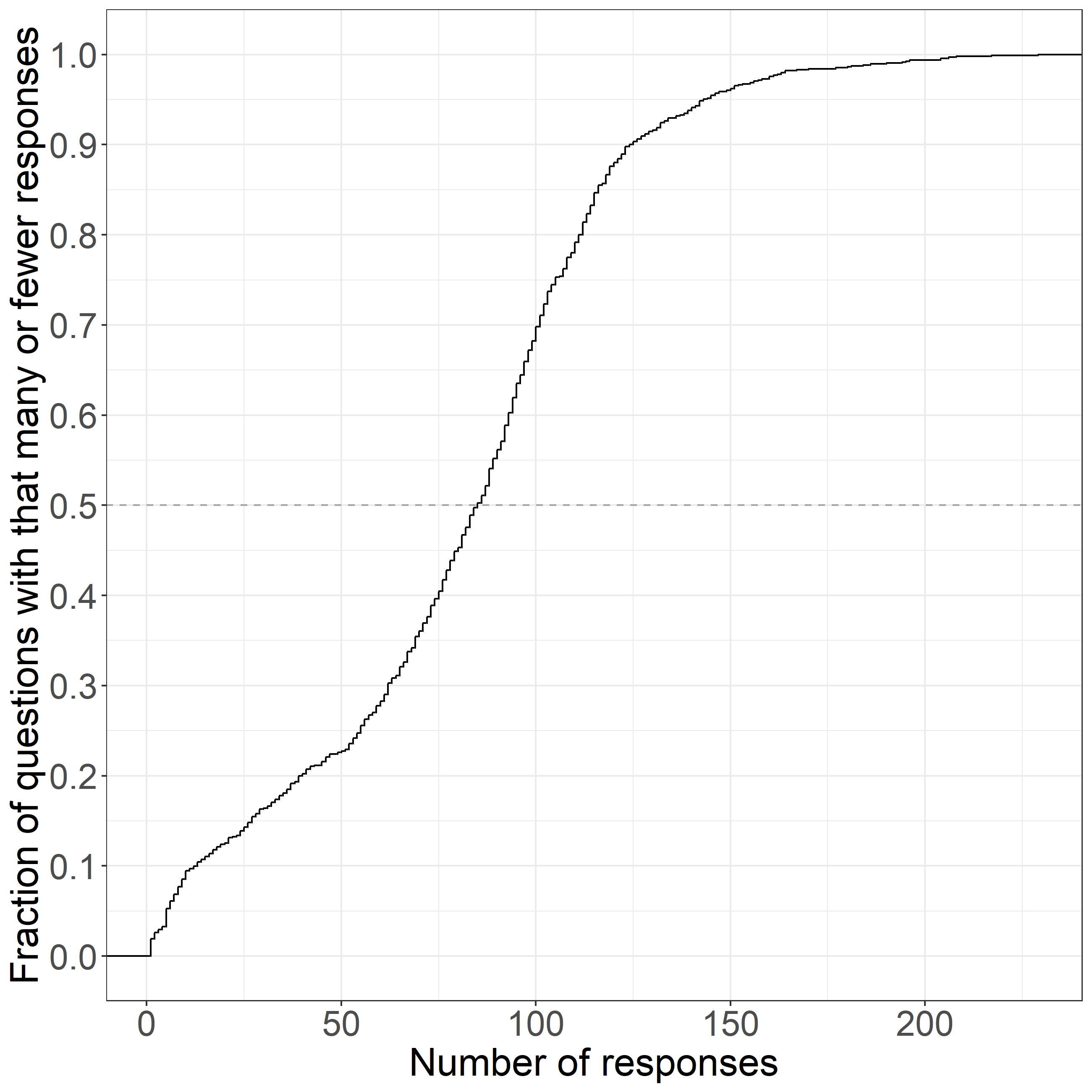}
    \caption{The cumulative distribution of responses to each question in Problem Roulette for Physics II. The dotted line corresponds to the median number of responses which is 85 for this data set.}
    \label{fig:ogive}
\end{figure}

We first matched all responses to Physics II questions in Problem Roulette to the university's enrollment records and removed responses from students who were not enrolled in the course and any responses that couldn't be matched to an enrollment record. We also removed responses that were blank and hence not marked as correct or incorrect by the system, which happens when the user is delivered a question and then exits the platform without answering it (presumably because they are done studying for the time being). Finally, we removed responses that were marked as ``retry" so that our data set only contained the student's first response to each question, excluding any responses provided after students received feedback and mimicking more typical testing environments in which students respond to each test item only once. We were left with 75,544 responses to 951 questions from 1,685 students, where students and questions were crossed given that each student could respond to each question. Because questions are served randomly to students, questions have different numbers of responses. In our data set, the most encountered question had 229 responses, the least encountered question had 1 response and the median number of responses was 85. Random effects modeling was used to account for this imbalance \cite{hoffman_longitudinal_2015, snijders_multilevel_2012} and is described later in this section. The cumulative distribution of responses is shown in Figure \ref{fig:ogive}.

To address our research questions, we then identified the CMC questions. We searched each question response for one of one hundred strings of text that could be indicative of a CMC question such as ``1 and 2 ", ``both b and d", or ``I, II, and III". Each question that included an answer pattern matching this list was then manually reviewed to see if it was in fact a CMC question according to our definition presented in Sec.\ref{sec:background}. For example, a question that asked students to rank the brightness of light bulbs in a circuit might contain the string ``I, II, and III" and hence be errantly flagged as a CMC question. Questions not marked as CMC were then marked as non-CMC questions. From this exercise, we identified 44 CMC questions and 907 non-CMC questions in our data set. 
For both types of questions, we summed the total number of correct first responses and divided by the number of first responses to get overall accuracy rates.

To determine how much of the performance difference between CMC questions and non-CMC questions could be explained by the differences in the type of knowledge students would need to answer the question, we sampled the non-CMC questions to estimate the percentage of non-CMC questions that were conceptual vs ``plug-and-chug". To do so, we categorized the first fifty questions in the Problem Roulette database as conceptual or ``plug-and-chug". If correctly answering the question required a numeric or symbolic calculation more than just a multiplicative factor (e.g. a proportional reasoning question), we marked the question as ``plug-and-chug" and otherwise conceptual. Examples of each type of question are included in the supplemental material. Afterward, we randomly sampled the remaining questions in the database until we had categorized 15\% of the non-CMC questions (136 questions). In doing so, we found that 37\% of our sampled non-CMC questions were conceptual while the remaining 63\% were ``plug-and-chug" (Table \ref{tab:c-p-table}).

\begin{table}[]
\caption{Percent of each question type by conceptual or ``plug-and-chug" categorization}
\label{tab:c-p-table}
\begin{tabular}{p{.30\linewidth} p{0.25\linewidth} p{0.25\linewidth}} \hline
 & CMC questions & Sampled non-CMC questions \\ \hline
Conceptual & 100 & 37 \\
``Plug-and-chug" & 0 & 63 \\ \hline
\end{tabular}
\end{table}

To account for prior preparation, we used a metric called grade point average in other courses (GPAO) as well as standardized test scores. GPAO is calculated in relation to a specific course and is generally calculated as the student's grade point average in all other courses the student has taken up to and including the semester the student took the specific course \cite{huberth_computer-tailored_2015}. For example, the GPAO for a student who enrolled in Fall 2019 and took Physics II in Winter 2021 would be their grade point average of all the courses they took between Fall 2019 and Winter 2021 inclusive except for Physics II. GPAO is therefore a measure of how well a student has performed relatively in their courses at their institution and has been found to be more predictive of a course grade than a student's high school GPA or their SAT or ACT scores \cite{koester_patterns_2016, matz_patterns_2017}. 

Students at the university were able to submit either ACT or SAT scores and thus, both are not recorded for most students. If a student only submitted SAT scores, we used ACT concordance tables to convert from the SAT to ACT score. We chose this conversion because SAT scores map uniquely to an ACT score while an ACT score maps to a range of SAT scores. If the student had both scores, the existing ACT score was used and the SAT score was not converted. Overall, only 58 (3.5\%) students had neither an ACT nor SAT score recorded in the student data warehouse.

We then split students into a high-performing group if their GPAO was at least 3.7 (an average grade of A- or better) and otherwise a lower-performing group and compared their overall accuracy rates on CMC and non-CMC questions. For the purpose of referring to the grades students earn in general (via GPAO), we use ``A'' to refer to any A-level (A-, A, A+). We discretized this continuous variable because we are interested in whether CMC questions disproportionately help high-performing students. For reference, grades at the university are based on a standard 4.0 scale (A is 4.0, A- is 3.7, B+ is 3.3, B is 3.0, etc.)

Likewise, to understand how CMC questions could differentiate students who would earn a high versus low grade in Physics II, we also categorized students as a high grade earner if they received at least an A- in Physics II and a lower grade earner if they earned a B+ or less; roughly 40\% of students were classified as high grade earners (see Figure \ref{fig:app_grade}). We base this choice on the observation that earning an ``A" is the most common grade and that the percentage of students earning other letter grades drops off steadily after. We note that B-level grades could also be included in the higher grade category. However, in our experience working with faculty, we have noticed that when faculty refer to students who earned a high grade in their course, they are nearly always referring to students who earned ``A"s. Further, including B-level grades would result in around 70\% of students being classified as high grade earners.

To understand how CMC questions may affect performance gaps, we compared performance on CMC questions to non-CMC questions for five demographic groups based on data in our student data warehouse. Across all five variables, students with missing data were only excluded from the analysis for the specific demographic variable that they were missing.

To better understand the impact of question style on accuracy, we looked at performance on CMC and non-CMC questions by individual students. We compared each student's fraction of correct responses to non-CMC and CMC questions. Due to the possibility that the randomly presented questions the student saw might have only included one or two CMC questions and thus result in an exaggerated performance on the questions, we required that the student had responded to at least five CMC questions to be included in this portion of the analysis. 

Finally, to gain further understanding of any effects observed, we ran a series of mixed effects logistic regression models. We first ran five ``simple" regression models, using one of each of the five demographic variables as a binary variable, question type (complex multiple-choice or non-complex multiple-choice), and an interaction term as inputs and whether the question was answered correctly (yes or no) as an output. For each of these models, we included the term the student was enrolled, the question ID and the student as random intercepts to account for possible baseline performance differences based on the term, the individual question, and the individual student. We also included a random slope based on the student and question type because the relationship between performance on CMC and non-CMC questions might not be the same for all students. If there was a differential effect between the demographic or student groups, the interaction term would be statistically significant. We also ran another two more ``simple" models, one with GPAO (as a continuous variable) and one with ACT score, question type, an interaction term, and the same random intercepts and slopes as before.

Finally, to account for how some of the variables may be measuring similar effects, we ran a more complex logistic regression model including all five demographic variables, GPAO, ACT score, and all interactions between these variables and question type. We also included the number of CMC questions the student attempted as a covariate in the model as well as an interaction term between this and question type to take into account that students might simply be less familiar with CMC questions thus have worse performance. This more complex model also included the random slope and intercepts. If an effect is statistically significant in most of these models, we can be more confident that what we are measuring is an actual signal in the data and not just noise. We performed all analysis in R \cite{r_core_team_r_2021} and used the \texttt{lme4} package to run the nested models \cite{bates_fitting_2015}. As in Young and Caballero \cite{young_nicholas_t_predictive_2021}, the percentage of responses attributed to each variable in our models are shown in Table \ref{tab:distribution}. 

Across these models, continuous variables were scaled and centered as appropriate. We centered GPAO and ACT scores based on the average GPAO and ACT among the students in the data set as a GPAO or ACT of zero is highly unlikely to occur in practice. Under this transformation, the intercept corresponds to a student with an average GPAO and average ACT score in this sample (3.49 and 32.75, respectively). Because it is possible for a student to never answer a CMC question, the number of CMC questions attempted was not centered. However, due to the wide range of CMC questions students saw, we scaled this variable by the standard deviation of the sample (4.21). For students who attempted more than 44 CMC questions (i.e., were randomly served some CMC questions multiple times), we replaced their actual number of CMC questions attempted with 44. This was rare, corresponding to only 0.2\% of the sample.

In addition, any cases with missing data were removed. For these models, there were 67,279 usable responses in 950 questions and 1,493 students, where questions and students are crossed, across 13 terms.

\begin{table}[]
\caption{Percent of responses to questions by variables in the regression models by category as well as the amount of missing data.}
\label{tab:distribution}
\begin{tabular}{p{.5\linewidth}p{0.2\linewidth}p{0.15\linewidth}} \hline
Variable & Split & Missing      \\ \hline
\begin{tabular}[c]{@{}l@{}}Question type\\ (non-CMC \textbar\; CMC)\end{tabular}  & 95/5  & 0            \\
\begin{tabular}[c]{@{}l@{}}Sex\\ (female \textbar\; male)\end{tabular}  & 30/70 & \textless{}1 \\
\begin{tabular}[c]{@{}l@{}}Race/ethnicity\\ (B/H/M/N \textbar\; A/w)\end{tabular}  & 13/80 & 7            \\
\begin{tabular}[c]{@{}l@{}}Parental education\\ (first-gen \textbar\; cont.-gen )\end{tabular} & 10/89  & 2            \\
\begin{tabular}[c]{@{}l@{}}Socioeconomic status\\ (low \textbar\; medium/high)\end{tabular}   & 48/52 & \textless{}1 \\
\begin{tabular}[c]{@{}l@{}}Residency\\ (international \textbar\; domestic)\end{tabular}   & 6/94  & \textless{}1 \\ \hline
\end{tabular}
\end{table}

\section{Results}\label{sec:results}
\subsection{Descriptive statistics}
Examining overall performance, we found that students performed worse on CMC questions compared to non-CMC questions. Students submitted 3,143 responses to CMC questions with an accuracy rate of 42.0\% (95\% confidence interval: 40.3\%, 43.7\%) compared to 72,401 responses to non-CMC questions with an accuracy rate of 49.9\% (49.6\%, 50.3\%). That is, the rate at which students correctly answered CMC questions was 8.0 (6.2, 9.7) percentage points lower than that at which they correctly answered non-CMC questions.

\begin{figure*}
    \centering
    \includegraphics[width=.85\linewidth]{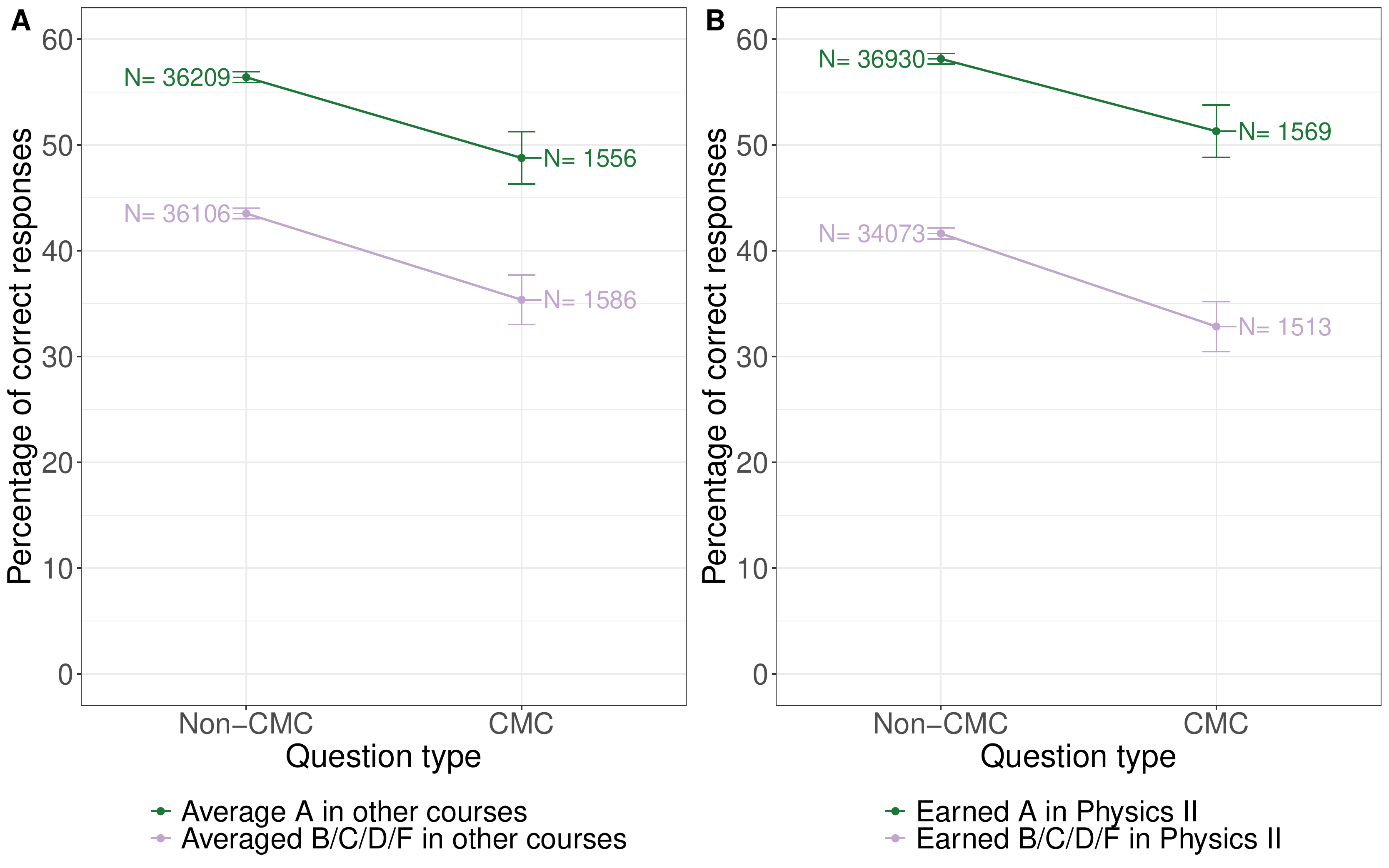}
    \caption{Percent of CMC and non-CMC questions answered correctly based on the average grade students earned in their other classes before or while enrolled in Physics II (panel A) and grade earned in Physics II (panel B). 
    Numbers above each data point are the number of responses from that grade group to that type of question and the error bars are 95\% confidence intervals. Both groups of students do equally worse on CMC questions. }
    \label{fig:gpao_grade_CMC}
\end{figure*}

To understand whether this performance difference was due to the format or content of the question (conceptual vs. numerical), we compared student accuracy on a subset of non-CMC questions. We found that 50.6\% of responses to ``plug-and-chug" questions were correct (95\% confidence interval: 49.4\%, 51.8\%) while 47.6\% of responses to conceptual questions were correct (95\% CI: 46.2\%, 49.1\%).
That is, students perform 5.63 (95\% CI 3.36, 7.90) percentage points better on non-CMC conceptual questions than they do on CMC conceptual questions.

The results of taking prior performance and grades earned in Physics II into account when answering CMC and non-CMC questions are shown in Figure \ref{fig:gpao_grade_CMC}. Looking at prior performance first (Fig. \ref{fig:gpao_grade_CMC}A), we notice that regardless of a student's grades in their other courses, students correctly answered non-CMC questions at a higher rate than CMC questions. However, students who typically earn ``A"s in their other courses correctly answered CMC questions at a higher rate than students who typically earned less than ``A"s in their other courses. Looking at the discrepancy in performance on the question types between students who earned ``A"s in their other courses and those who did not, we find the gap to be about the same size as previously described---around 8 percentage points.

Looking specifically at the grades students earned in Physics II, we find a similar result (Fig. \ref{fig:gpao_grade_CMC}B). Students who earned ``A"s in Physics II correctly responded to non-CMC questions more often than they did CMC questions and correctly answered CMC questions more often than students who didn't earn ``A"s in Physics II correctly responded to non-CMC questions. Looking at the difference in performance between the two question types based on the student's Physics II grade, we notice that the gap is slightly smaller for students who earned ``A"s compared to those who did not (6.8 vs. 8.8 points).

\begin{figure}
    \centering
    \includegraphics[width=.85\linewidth]{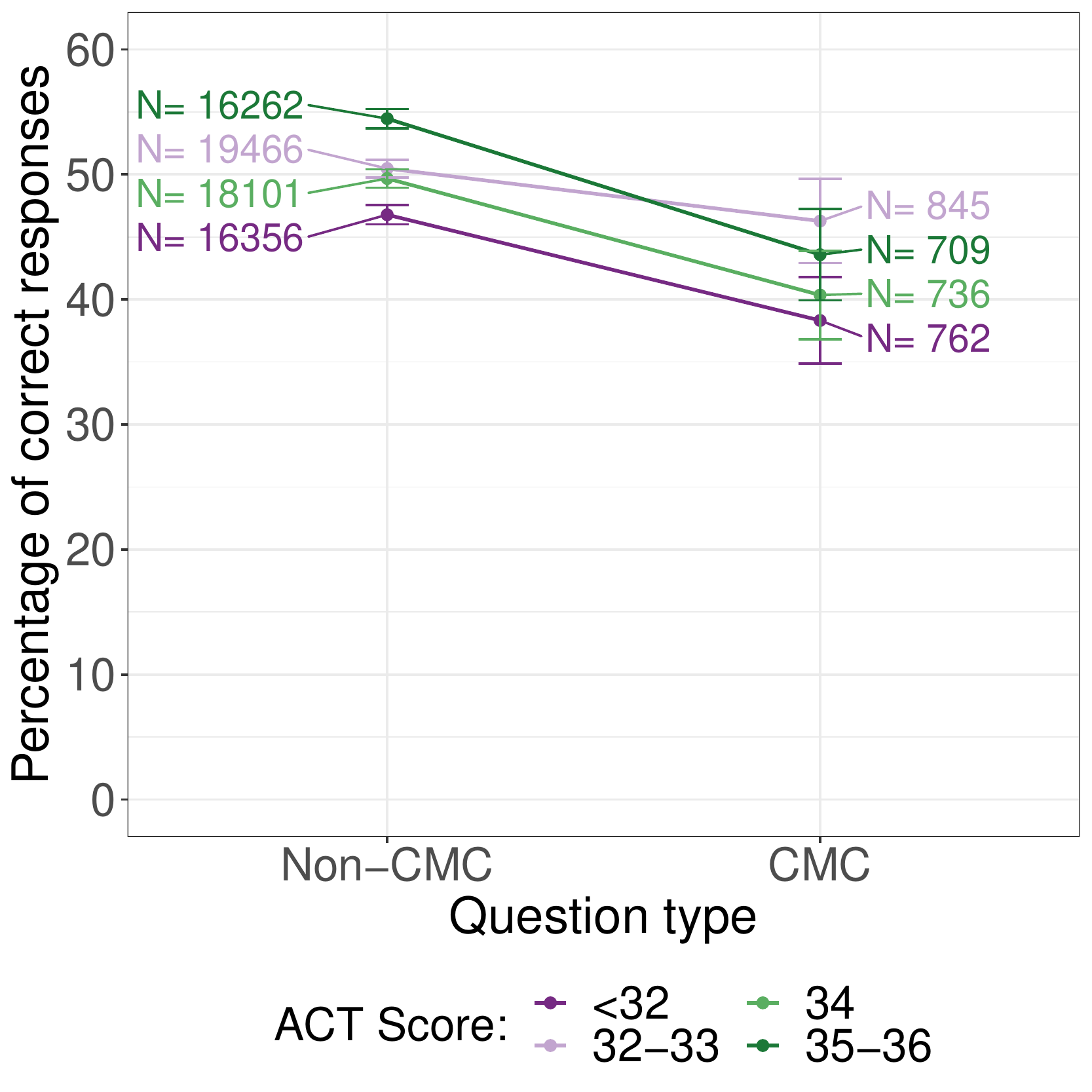}
    \caption{Percent of CMC and non-CMC questions answered correctly based on student's ACT score. Higher ACT scores are represented in green while lower ACT scores are represented in purple. Each category corresponds to around a quarter of students in our data set. Numbers above each data point refer to the number of responses from that group to that type of question and the error bars are 95\% confidence intervals. Students tend to do better on non-CMC questions compared to CMC questions}
    \label{fig:act_CMC}
\end{figure}

We then considered prior performance through the lens of standardized test scores. The result is shown in Figure \ref{fig:act_CMC} where the ACT score groups were chosen to correspond to roughly a quarter of students. We find that regardless of ACT score, students perform better on non-CMC questions. Notably, while performance on non-CMC questions tends to increase with ACT score, performance on CMC questions is largely unchanged.

Next, we examined performance on the CMC and non-CMC questions from the perspective of differentiating students within a course. That is, given a student who answered a CMC or non-CMC question correctly, what was the probability that the student would earn an ``A" in Physics II? If CMC questions are better at discriminating between ``A" students and other students, we would expect that answering CMC questions correctly would lead to a higher probability of earning an ``A".  However, we find that there is no apparent relationship between students' rate of accuracy on CMC and non-CMC questions and their probability of earning an ``A" in Physics II (Table \ref{tab:A_prob}).

\begin{table}[]
    \centering
    \caption{Probability and 95\% confidence intervals that a student who answered each type of multiple-choice question correctly would go on to earn an ``A" in Physics II.}
    \begin{tabular}{p{.35\linewidth}p{.35\linewidth}p{.25\linewidth}} \hline
        Type of Question & Probability & Total Number of Responses \\ \hline
         CMC Questions & 0.618 (0.585, 0.652) & 1,302 \\
         Non-CMC Questions & 0.602 (0.596, 0.609) & 35,660 \\ \hline
    \end{tabular}
    \label{tab:A_prob}
\end{table}

\begin{figure*}
    \centering
    \includegraphics[width=.95\linewidth]{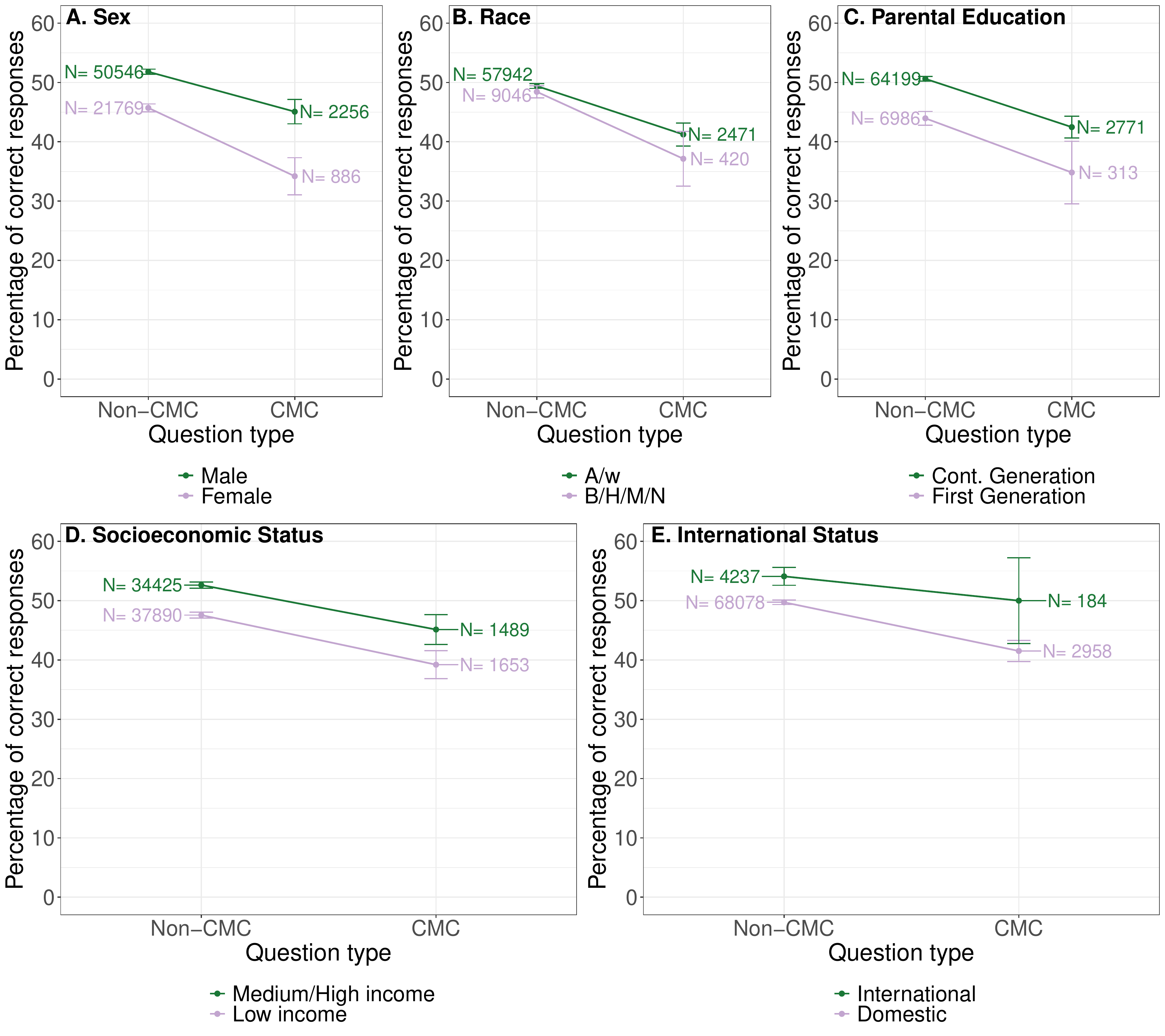}
    \caption{Comparison of the percentage of correct responses to CMC and non-CMC questions by Physics II students split by demographic groups. Error bars on each point represent 95\% confidence intervals and the number above each point is the number of responses by that demographic group to that question type. Larger plots in the second row represent the demographic categories found to have a statistically significant interaction with question type in our regression models.}
    \label{fig:demo_plot}
\end{figure*}

We then looked at how various demographic groups performed on CMC and non-CMC questions (Figure \ref{fig:demo_plot}). All groups performed better on the non-CMC questions but the size of the performance gap varied between the groups. For example, the CMC-non-CMC performance gap was smallest for international students (4\%) and largest for Black, Hispanic, Multiracial, and Native American students and female students (about 11\% for both groups). In general, we find that minoritized students have a larger performance gap between the two types of questions than their majoritized peers do. The difference in the performance gap varies between 0.9\% and 4.8\%, with the smallest corresponding to the difference in performance gap between low-income and medium/high-income students and the largest corresponding to the difference in performance between male and female students. Practically, this result means that if male students on average had a performance penalty of $\delta$ percentage points on CMC questions compared to non-CMC questions, a female student on average would have a performance gap of $\delta + 4.8$ percentage points.

\begin{figure}
    \centering
    \includegraphics[width=.95\linewidth]{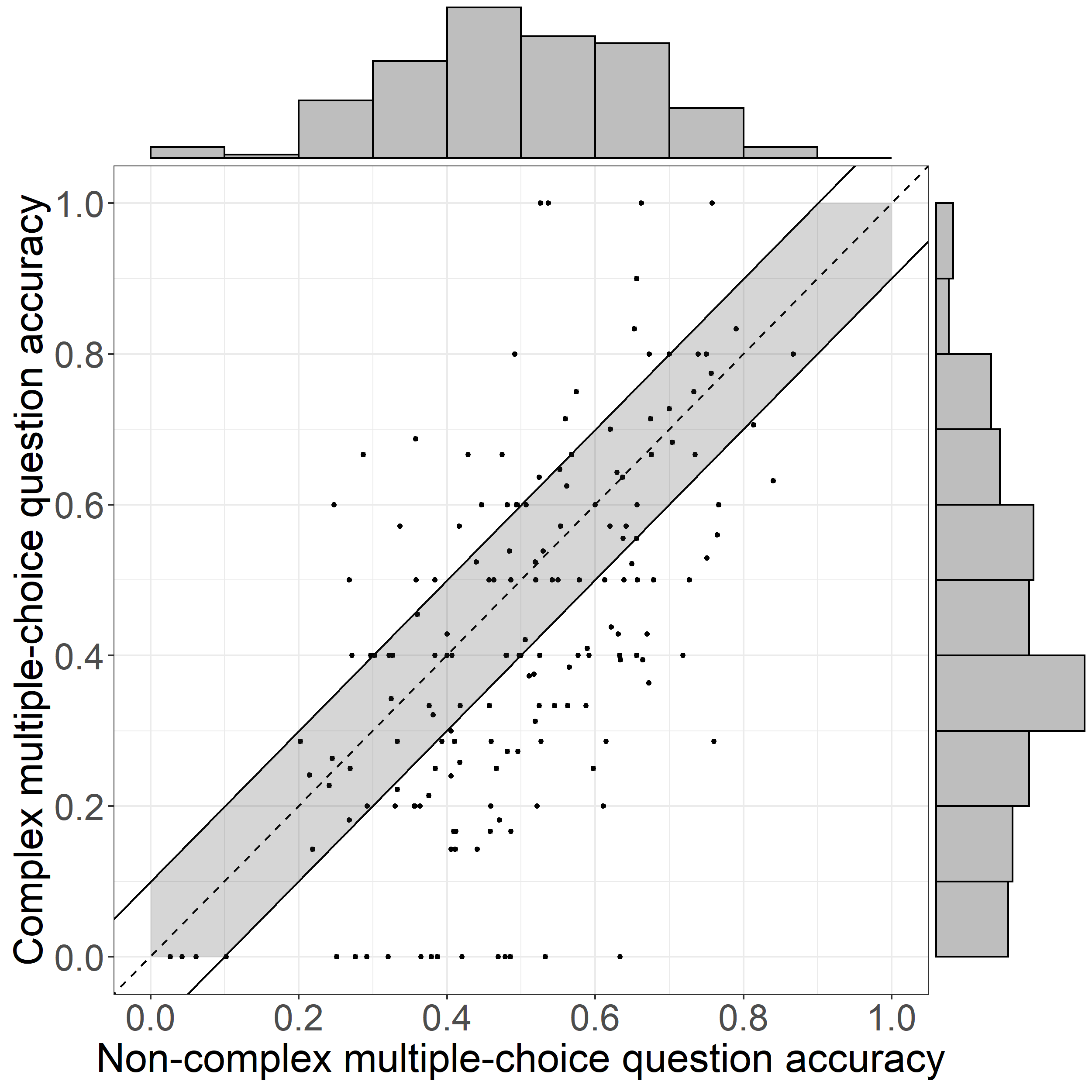}
    \caption{Comparison of student accuracy on non-CMC compared to CMC questions among students who answered at least five of each question type. The diagonal line denotes equal performance on the two types of questions. The gray region denotes where performance on the two types of questions are within ten percentage points of each other, or the equivalent of a letter grade on the standard scale where 90 to 100 equates to an A grade. The edge histograms show the relative number of students in each 10\% band of accuracy. In general, students are less accurate at answering CMC than non-CMC questions, shown by there being more dots below the line of parity.}
    \label{fig:mc_v_cmc}
\end{figure}

Finally, we compared individual student performance on the two types of questions (Figure \ref{fig:mc_v_cmc}). Of the students who saw at least five of each question type, most students (65.7\%) answered a higher fraction of the non-CMC questions correctly compared to the CMC questions. Requiring a practically significant difference of at least ten percentage points, the equivalent of a full letter grade, we find that 46\% of students performed better on the non-CMC questions, 16\% performed better on the CMC questions, and 38\% performed the same or within the 10 percentage point threshold (marked by the grey region in Figure \ref{fig:mc_v_cmc}).

\subsection{Regression models}

\begin{table*}[]
\caption{Odds ratios for each regression model for answering a question correctly; *** \textit{p} \textless{} .001, ** \textit{p} \textless{} .01, and * \textit{p} \textless{} .05.}
\label{tab:simp_models_2}
\begin{tabular}{p{.16\linewidth}p{.11\linewidth}p{.11\linewidth}p{.11\linewidth}p{.11\linewidth}p{.11\linewidth}p{.11\linewidth}p{.11\linewidth}} \hline
Variable & Model 1 & Model 2 & Model 3 & Model 4 & Model 5 & Model 6 & Model 7 \\ \hline
Intercept & \begin{tabular}[c]{@{}c@{}}0.78***\\ (0.69, 0.89)\end{tabular} & \begin{tabular}[c]{@{}c@{}}0.74***\\ (0.66, 0.83)\end{tabular} & \begin{tabular}[c]{@{}c@{}}0.75***\\ (0.67, 0.85)\end{tabular} & \begin{tabular}[c]{@{}c@{}}0.74***\\ (0.65, 0.85)\end{tabular} & \begin{tabular}[c]{@{}c@{}}0.72***\\ (0.64, 0.82)\end{tabular} & \begin{tabular}[c]{@{}c@{}}0.73***\\ (0.63, 0.85)\end{tabular} & \begin{tabular}[c]{@{}c@{}}0.75***\\ (0.67, 0.84)\end{tabular} \\
\begin{tabular}[c]{@{}c@{}}Question Type\\ (CMC =1)\end{tabular} & \begin{tabular}[c]{@{}c@{}}0.75*\\ (0.58, 0.97)\end{tabular} & \begin{tabular}[c]{@{}c@{}}0.74*\\ (0.57, 0.95)\end{tabular} & \begin{tabular}[c]{@{}c@{}}0.70**\\ (0.54, 0.90)\end{tabular} & \begin{tabular}[c]{@{}c@{}}0.69**\\ (0.53, 0.90)\end{tabular} & \begin{tabular}[c]{@{}c@{}}0.70**\\ (0.54, 0.89)\end{tabular} & \begin{tabular}[c]{@{}c@{}}0.71**\\ (0.55, 0.91)\end{tabular} & \begin{tabular}[c]{@{}c@{}}0.71**\\ (0.55, 0.91)\end{tabular} \\
\begin{tabular}[c]{@{}c@{}}Sex\\ (female = 1)\end{tabular} & \begin{tabular}[c]{@{}c@{}}0.81***\\ (0.73, 0.89)\end{tabular} &  &  &  &  &  &  \\
\begin{tabular}[c]{@{}c@{}}Race\\ (B/H/M/N = 1)\end{tabular} &  & \begin{tabular}[c]{@{}c@{}}0.92\\ (0.81, 1.05)\end{tabular} &  &  &  &  &  \\
\begin{tabular}[c]{@{}c@{}}Parental Education\\ (First gen = 1)\end{tabular} &  &  & \begin{tabular}[c]{@{}c@{}}0.78***\\ (0.67, 0.90)\end{tabular} &  &  &  &  \\
\begin{tabular}[c]{@{}c@{}}Socioeconomic Status\\ (Low income = 1)\end{tabular} &  &  &  & \begin{tabular}[c]{@{}c@{}}0.97\\ (0.88, 1.06)\end{tabular} &  &  &  \\
\begin{tabular}[c]{@{}c@{}}International Status\\ (International = 1)\end{tabular} &  &  &  &  & \begin{tabular}[c]{@{}c@{}}1.22*\\ (1.00, 1.47)\end{tabular} &  &  \\
\begin{tabular}[c]{@{}c@{}}GPAO\\ (centered at 3.492)\end{tabular} &  &  &  &  &  & \begin{tabular}[c]{@{}c@{}}2.16***\\ (1.96, 2.39)\end{tabular} &  \\
\begin{tabular}[c]{@{}c@{}}ACT score\\ (centered at 32.75)\end{tabular} &  &  &  &  &  &  & \begin{tabular}[c]{@{}c@{}}1.08***\\ (1.07, 1.10)\end{tabular} \\
Sex * Question Type & \begin{tabular}[c]{@{}c@{}}0.81*\\ (0.67, 0.99)\end{tabular} &  &  &  &  &  &  \\
Race * Question Type &  & \begin{tabular}[c]{@{}c@{}}0.82\\ (0.63, 1.07)\end{tabular} &  &  &  &  &  \\
Parental Education * Question Type &  &  & \begin{tabular}[c]{@{}c@{}}0.98\\ (0.74, 1.30)\end{tabular} &  &  &  &  \\
Socioeconomic Status * Question Type &  &  &  & \begin{tabular}[c]{@{}c@{}}1.04\\ (0.87, 1.24)\end{tabular} &  &  &  \\
International Status * Question Type &  &  &  &  & \begin{tabular}[c]{@{}c@{}}1.30\\ (0.92, 1.84)\end{tabular} &  &  \\
GPAO *Question Type &  &  &  &  &  & \begin{tabular}[c]{@{}c@{}}1.06\\ (0.85, 1.32)\end{tabular} &  \\
ACT Score * Question Type &  &  &  &  &  &  & \begin{tabular}[c]{@{}c@{}}0.98\\ (0.95, 1.01)\end{tabular}   \\ \hline
\end{tabular}
\end{table*}

To assess if those differences in performance gaps were significant, we developed regression models. The results of the ``simple" regression models are shown in Table \ref{tab:simp_models_2}. In models 1-7, we see that question type is always significant with an odds ratio ranging between 0.70 and 0.75, meaning that the odds of a student answering a question correctly decrease by between 0.691 and 0.753 when the question is a CMC question instead of a non-CMC question. Equivalently, the odds of a student answering a question correctly are 1.33 and 1.45 times higher when the question is a non-CMC question instead of a CMC question.

We also find that many of the demographic categories (sex, first-generation status, and international student status) are statistically significant, meaning that there are differing accuracy rates based on some of the student's demographic characteristics. Of course, the demographic results should not be interpreted as representing cause-and-effect relationships. That is, being a first-generation student doesn't cause lower performance on Problem Roulette questions; rather, first-generation students perform less well on Problem Roulette questions than continuing-generation students.

Considering the measures of prior preparation, we find that both GPAO and ACT scores are statistically significant and have odds ratios greater than one. This means that students with higher GPAOs and ACT scores are associated with greater odds of getting a question correct.

Based on the interaction terms, we see that the sex and CMC interaction is statistically significant. That is, our model finds that female students are predicted to answer CMC questions correctly disproportionately less often than male students as compared to non-CMC questions. We find no evidence that that is the case for any of the other demographic groups.



\begin{table}[]
\caption{Odds ratios for mixed effects model with students and questions as crossed random effects; *** \textit{p} \textless{} .001, ** \textit{p} \textless{} .01, * \textit{p} \textless{} .05, $\cdot$ \textit{p} \textless{} .10.}
\label{tab:full_model_2}
\begin{tabular}{p{.50\linewidth}p{.30\linewidth}} \hline
Variable  & Odds Ratio  \\ \hline
Intercept & \begin{tabular}[c]{@{}c@{}}0.79**\\ (0.68, 0.91)\end{tabular} \\
\begin{tabular}[c]{@{}c@{}}Question Type\\ (CMC = 1)\end{tabular} & \begin{tabular}[c]{@{}c@{}}0.74*\\ (0.55, 0.99)\end{tabular} \\
\begin{tabular}[c]{@{}c@{}}Sex\\ (female = 1)\end{tabular} & \begin{tabular}[c]{@{}c@{}}0.80***\\ (0.73, 0.88)\end{tabular} \\
\begin{tabular}[c]{@{}c@{}}Race\\ (B/H/M/N = 1)\end{tabular} & \begin{tabular}[c]{@{}c@{}}1.07\\ (0.95, 1.20)\end{tabular} \\
\begin{tabular}[c]{@{}c@{}}Parental Education\\ (First gen = 1)\end{tabular} & \begin{tabular}[c]{@{}c@{}}1.02\\ (0.89, 1.17)\end{tabular} \\
\begin{tabular}[c]{@{}c@{}}Socioeconomic Status\\ (Low income = 1)\end{tabular} & \begin{tabular}[c]{@{}c@{}}0.95\\ (0.87, 1.04)\end{tabular} \\
\begin{tabular}[c]{@{}c@{}}International Status\\ (International = 1)\end{tabular} & \begin{tabular}[c]{@{}c@{}}1.11\\ (0.92, 1.34)\end{tabular} \\
\begin{tabular}[c]{@{}c@{}}GPAO\\ (centered at 3.492)\end{tabular} & \begin{tabular}[c]{@{}c@{}}1.95***\\ (1.75, 2.17)\end{tabular} \\
\begin{tabular}[c]{@{}c@{}}ACT\\ (centered at 32.75)\end{tabular} & \begin{tabular}[c]{@{}c@{}}1.04***\\ (1.02, 1.06)\end{tabular} \\
\begin{tabular}[c]{@{}c@{}}CMC-questions seen\\ (scaled by standard deviation)\end{tabular} & \begin{tabular}[c]{@{}c@{}}1.05**\\ (1.02, 1.09)\end{tabular} \\
Sex * Question Type & \begin{tabular}[c]{@{}c@{}}0.85 $\cdot$\\ (0.71, 1.03)\end{tabular} \\
Race * Question Type & \begin{tabular}[c]{@{}c@{}}0.79 $\cdot$\\ (0.60, 1.02)\end{tabular} \\
Parental Education * Question Type & \begin{tabular}[c]{@{}c@{}}1.12\\ (0.83, 1.51)\end{tabular} \\
Socioeconomic Status * Question Type & \begin{tabular}[c]{@{}c@{}}1.22*\\ (1.01, 1.46)\end{tabular} \\
International Status * Question Type & \begin{tabular}[c]{@{}c@{}}1.43*\\ (1.00, 2.05)\end{tabular} \\
GPAO * Question Type & \begin{tabular}[c]{@{}c@{}}1.24\\ (0.96, 1.60)\end{tabular} \\
ACT * Question Type & \begin{tabular}[c]{@{}c@{}}0.96*\\ (0.93, 1.00)\end{tabular} \\
CMC-questions seen * Question Type & \begin{tabular}[c]{@{}c@{}}0.96*\\ (0.94, 0.99)\end{tabular} \\ \hline 
\end{tabular}
\end{table}


Finally, we created a single model with all demographic terms, prior preparation measures, and interactions along with the number of CMC questions attempted (Table \ref{tab:full_model_2}). After controlling for all demographics and prior preparation, question type has an odds ratio statistically different from one, meaning that the odds of answering a CMC question correctly are lower than that of answering a non-CMC question correctly.

Second, the demographic variable of sex has an odds ratio statistically different from one, with male students having higher odds of answering a question correctly than female students. We also, unsurprisingly, find that both measures of prior preparation have odds ratios statistically greater than one, meaning that higher grades in college courses and higher ACT scores are associated with increased accuracy on Problem Roulette questions.

Third, we find that encountering more CMC questions (and by extension, trying more problems in Problem Roulette) is associated with higher odds of answering questions correctly.

Finally, examining the interaction terms, we find mixed results. The interactions between socioeconomic status and question type and between international status and question type are statistically different from one while the interactions between sex and question type and between race and question type are marginally significant.  The interaction between first-generation status and question type was not statistically significant.

\section{Discussion}\label{sec:discussion}
Here, we discuss each research question individually and then the limitations of the study overall.

\subsection{Individual Research Questions}\label{sec:RQs}
\subsubsection{How do students' performance on CMC questions compare to their performance on non-CMC questions in introductory physics?}

We found that students performed 8.0 percentage points worse on CMC questions compared to non-CMC questions. We also found from the regression models that question type was statistically significant, meaning that students performed differently on these types of questions. Taken together, our results suggest that, in line with various other studies (mainly conducted outside of physics), CMC questions are more difficult than non-CMC questions for students.

Second, we find evidence that some of the performance difference could also be explained by students lack of familiarity with CMC questions. That is, the odds ratio for number of CMC questions attempted in our regression model was statistically different from one. We note however, that because questions are served randomly from the system, a higher number of CMC questions attempted likely means that the student tried more questions overall and that extra practice could be what is responsible for the odds ratio being different than one. Unfortunately, the structure of our data does not allow us to separate these two effects.

\subsubsection{How does the type of knowledge students need to answer a question (conceptual vs. numerical) account for performance differences between CMC questions and non-CMC questions?}

When we considered that conceptual questions might be more difficult than numerical questions and that could explain the observed difference, we did not find sufficient evidence that that was the case. We found that on non-CMC questions, student accuracy on numerical questions was 3.0 percentage points higher than it was on conceptual questions. As the 95\% confidence intervals of the CMC/non-CMC performance difference and the numerical non-CMC/conceptual non-CMC performance differences do not overlap, these results suggest that the observed performance gap between CMC and non-CMC questions is not entirely due to CMC questions only being conceptual while the non-CMC questions were conceptual and ``plug-and-chug" questions. In addition, that students performed worse on CMC questions than non-CMC conceptual questions further supports this conclusion.

\subsubsection{What is the relationship between students' prior preparation and their ability to answer CMC questions correctly}
In terms of prior preparation, we find that ``A" students answer CMC questions correctly at a higher rate than non-``A" students do. However, the difference in performance between CMC and non-CMC questions is around 8 percentage points regardless of whether the student is an ``A" student in their other courses or not.

From the regression models, we find that a higher GPAO is associated with increased odds of answering a question correctly. However, the interaction term was not statistically different from one, suggesting that there is no disproportionate difference in performance on CMC and non-CMC questions.

When considering prior performance as measured by ACT scores, we find a different story. Not only is a higher ACT score associated with increased odds of answering a question correctly but there is also an interaction effect between ACT score and question type. However, the results can likely be explained as students with higher ACT scores do better on non-CMC questions than students with lower ACT scores, but everyone performs equally badly on CMC questions (Fig \ref{fig:act_CMC}).


While we cannot address the potential for clueing directly, our results suggest that if clueing is occurring, it is not helping lower-ability students perform better on CMC questions by eliminating primary answer options due to the available secondary choices as hypothesized by Albanese \cite{albanese_type_1993}. If clueing were helping lower-ability students, we would expect the performance gap between CMC and non-CMC questions to be smaller for non-``A" students than for ``A" students or for students with lower versus higher ACT scores because clueing would not help on the non-CMC questions. In addition, clueing would only benefit the non-``A" students and students with lower ACT scores and would be of little benefit to ``A" students or higher ACT scores who were assumed to know the material. Instead, we found the performance gap to be similar.


\subsubsection{How well can a CMC question differentiate between students who go on to earn a high versus low grade in their physics course?}
We find that CMC questions are able to differentiate between students who went on to earn ``A"s in Physics II from those who did not. However, we do not find that CMC questions offer any practical differentiation benefits from non-CMC questions as the difference in performance gaps was 2 percentage points. Therefore, our data does not support using CMC questions instead of non-CMC as a way to differentiate top-performing students from everyone else.

When considering the probability that someone who answered a CMC question correctly would go on to earn an ``A", we find a similar result. While the probability that someone who answered a CMC question correctly earns an ``A" in Physics II is higher by approximately 1.5 percentage points than the probability that someone who answered a non-CMC question correctly earns an ``A" in Physics II, the result is not statistically or practically different.

\subsubsection{How might CMC questions affect performance gaps between students of various demographic groups?}
We find that all demographic groups included in our study did worse on CMC than non-CMC questions. We also find that the differences between performance on CMC and non-CMC questions varied between demographic groups with international students having the smallest performance gap and Black, Hispanic, Multiracial, and Native American students having the largest performance gap. 

Based on the individual regression models, we found evidence that some majoritized demographic groups had higher odds of answering questions correctly than minoritized groups. However, only the sex and question type interaction had an odds ratio statistically different than one.

The full model that included all effects shows that the interactions between socioeconomic status and question type and between international status and question type were statistically significant, suggesting a disparate impact of CMC questions on these groups. We also found that the interactions between sex and question type and between race and question type were marginally significant. As the CMC penalty was largest for female students and Black, Hispanic, Multiracial, and Native American students, we interpret these marginally significant results as reflecting a lack of statistical power.

\subsection{Limitations}\label{sec:limits}
Our study has a few limitations that might affect the generalizability and applicability of our results.

First, our study did not occur in an actual testing environment and hence, might not necessarily hold there. While the questions in Problem Roulette are provided directly by course instructors and all questions are from prior Physics II midterms and finals, students may respond to questions differently based on if they are completed in a practice versus high-stakes examination format. In addition, research with multiple response multiple-choice questions, where students select all of the correct options for a multiple-choice test rather than a pick from a subset of choices containing pairs of responses, suggests that students might select more answer options for questions presented online compared to on paper \cite{olsho_online_2023}. While we are unable to determine if this is true for CMC questions as well based on our data, such an effect could also limit the applicability of our results to in-class, points-based assessments.

Second, our study did not directly compare CMC and non-CMC questions; that is, question content could be a confounding variable. Instead, our results are based on comparing overall performance on CMC and non-CMC questions. However, our random effects approach largely takes care of this from a modeling perspective. From a practical perspective, conducting a study with analogous questions presented in CMC and non-CMC formats might not be useful. For a CMC question to be presented as an analogous traditional multiple-choice question, there could only be one correct answer choice among the primary responses. Therefore, any question that could be posed as both a CMC and non-CMC question is not making particularly good use of the CMC format.

In addition, because Problem Roulette serves questions randomly, the impact of familiarity with CMC questions (as measured by the number of CMC questions attempted) cannot be separated from the impact of simply attempting more questions, regardless of the format. 

Finally, this study was conducted at an institution where the student population is a particular subset of the overall physics-taking population \cite{kanim_demographics_2020}. As such, we should not necessarily assume that the results would be applicable to institutions with greater representations of Black, Hispanic, Multi-racial, and Native American students or low-income students for example. In addition, the data set used in this study is based on students who opt-in to the Problem Roulette service, adding an additional confounding factor of how representative our study population is of the overall introductory physics-taking population. A comparison of students who used Problem Roulette compared to students who did not is included in the appendix and suggests that our sample is similar to the overall course population based on both grades earned in Physics II and in students` other courses, their ACT scores, and the demographics of the sample. We expect that the grades in Physics II should be slightly higher for students who used Problem Roulette because prior work has shown that regularly using Problem Roulette does result in a final grade boost \cite{black_quantifying_2023}.

\section{Future Work}\label{sec:future}
Complex multiple-choice questions are only one variant of alternatives to traditional multiple-choice questions; therefore, future work could examine the affordances and limitations that each provides in the context of physics. For example, multiple-response multiple-choice questions have been previously used in physics as an alternative to free-response questions \cite{wilcox_coupled_2014, wilcox_validation_2015} but work in other fields suggests this type of question is even more difficult than CMC questions for students \cite{frisbie_multiple_1992}. Alternatively, multiple-true-false questions strike a balance between the benefits of multiple responses while not being as difficult as CMC questions \cite{frisbie_multiple_1992}. Other studies suggest that students may actually prefer multiple-true-false questions compared to traditional multiple-choice questions \cite{frisbie_relative_1982, mobalegh_multiple_2012}.

To ensure that these studies are broadly applicable, research should be conducted at a variety of institutions under various conditions to simulate how students might engage with the material. These studies could include both in-class and online formats and in both low- and high-stakes settings.

Further, future work should examine CMC questions in disciplines other than physics. As introductory physics is considered difficult \cite{ornek_what_2008}, it is possible that the CMC penalty observed here might be underestimated for other disciplines where students did not find the material as difficult (or, alternatively, where low grades overall aren't as common).

\section{Conclusions}\label{sec:conclusions}
In this study, we found that students perform worse on CMC than non-CMC questions. We found this to be true for all demographic groups we investigated. In addition, we found that CMC questions disproportionately harm low-income and domestic students compared to medium and high income students and international students. Therefore, using CMC questions in a physics assessment can be an equity issue in the sense that some students might be penalized more than others.

From a practical lens, our work does not find many benefits to using these questions compared to others. For example, one of the main arguments for using CMC questions, that they are able to differentiate top students from lower-performing students, was not substantiated in our data. Therefore, in line with many others, we recommend against using this question format and encourage educators and researchers to explore other formats such as multiple-true-false and multiple-response multiple-choice questions in cases where forced-choice assessments are necessary.

\begin{acknowledgments}
We thank Erin Murray for her assistance in accessing the data for this study. Eric F. Bell was partly supported by the National Science Foundation through grant NSF-AST 2007065
\end{acknowledgments}

\section*{Appendix}\label{sec:appendix}

\begin{figure}
    \centering
    \includegraphics[width=.95\linewidth]{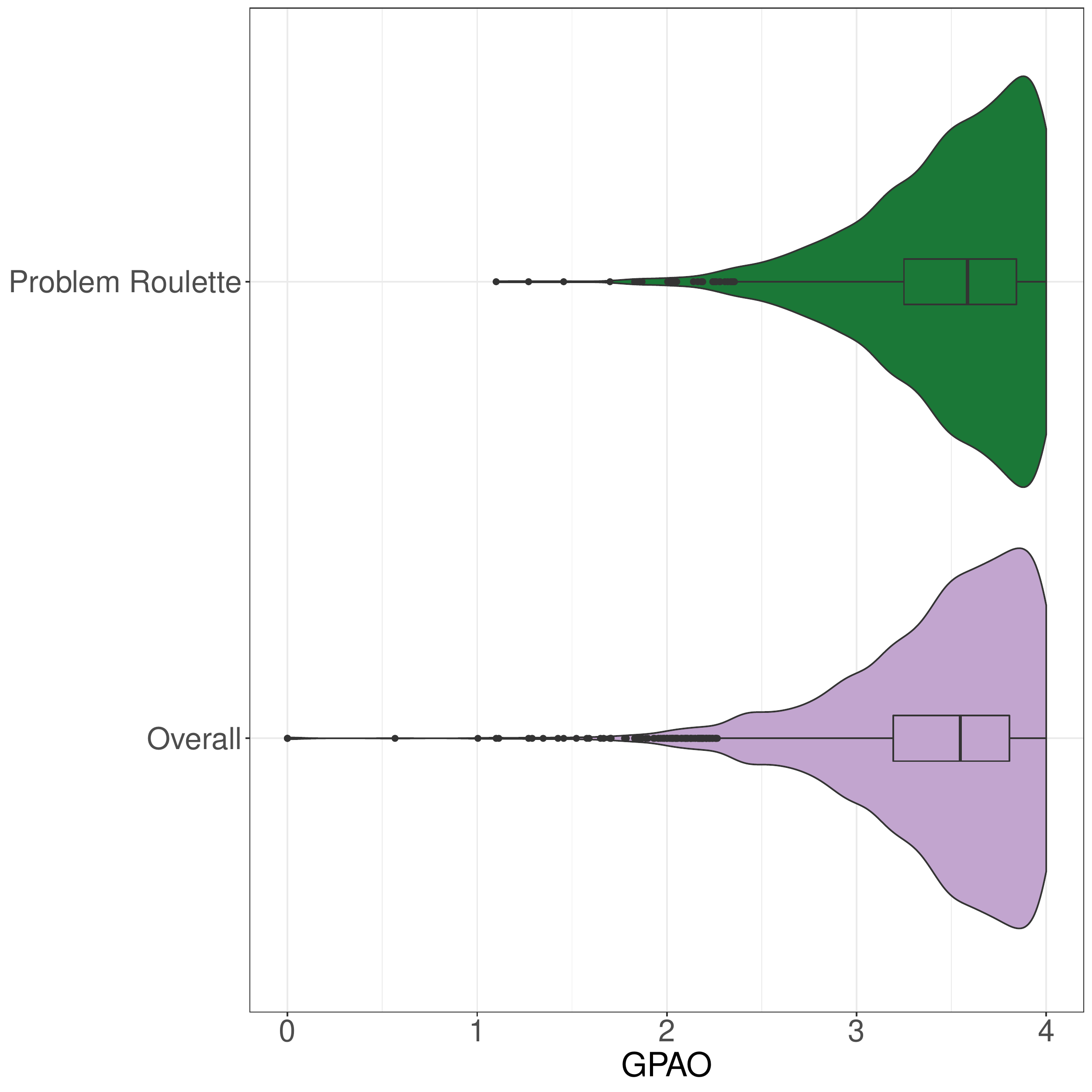}
    \caption{Comparison of the distribution of students who earned each grade (on average) in their other courses for students who used our Problem Roulette system and for all students in the course.}
    \label{fig:app_gpao}
\end{figure}

\begin{figure}
    \centering
    \includegraphics[width=.95\linewidth]{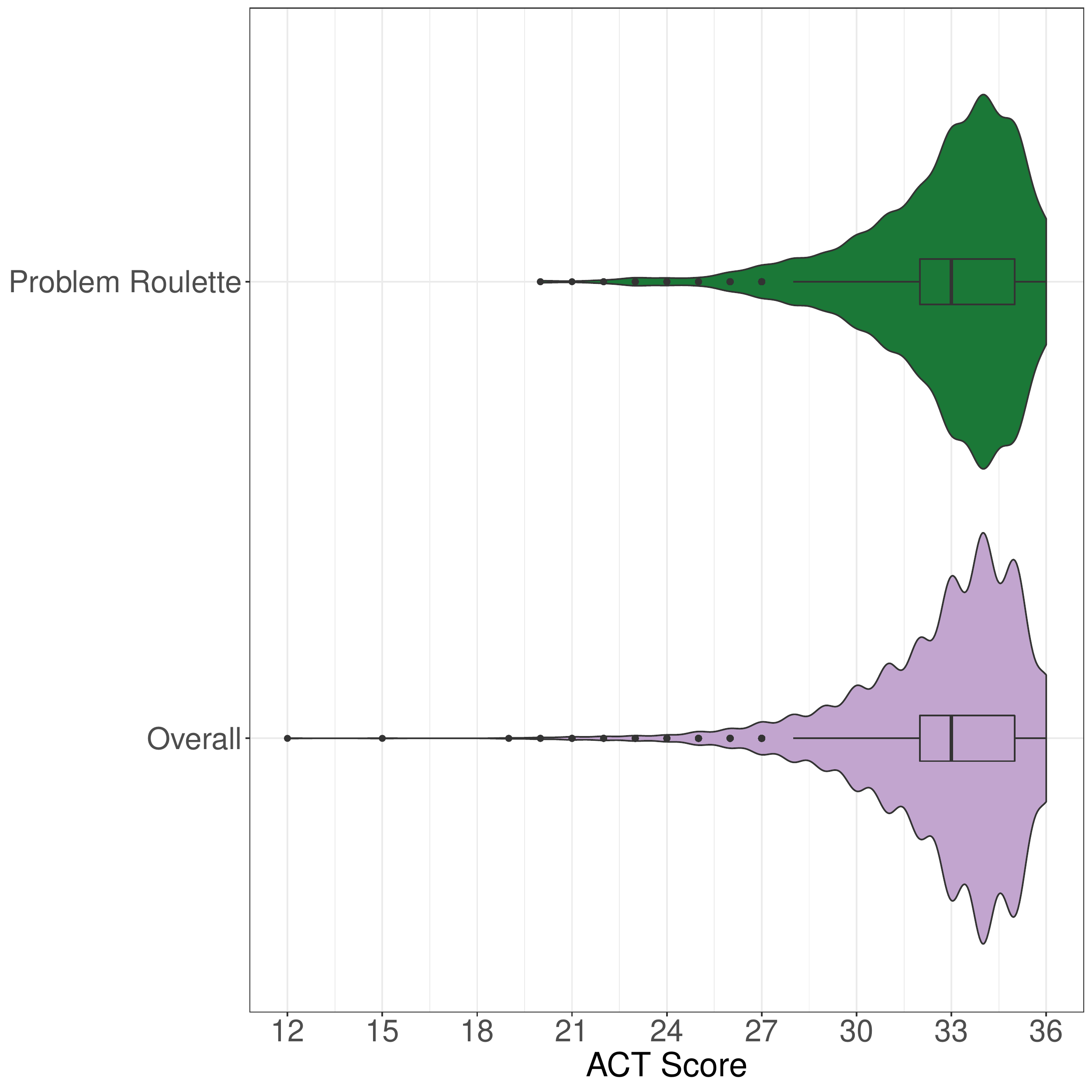}
    \caption{Comparison of the distribution of students who earned each ACT score for students who used our Problem Roulette system and for all students in the course.}
    \label{fig:app_act}
\end{figure}

Here, we compare Problem Roulette users in Physics II to everyone who took Physics II during the study period to provide additional context to our results.

First, we compared the grades students typically earned in their other classes for both Problem Roulette and non-Problem Roulette users (Fig. \ref{fig:app_gpao}). We find that overall, the distributions are similar; however, the median student who used Problem Roulette had a slightly higher GPAO compared to students in the course overall.

Second, we compared the ACT scores of students in the two groups (Fig. \ref{fig:app_act}). Again, we notice that the distributions are statistically similar for both groups of students.

\begin{figure}
    \centering
    \includegraphics[width=.95\linewidth]{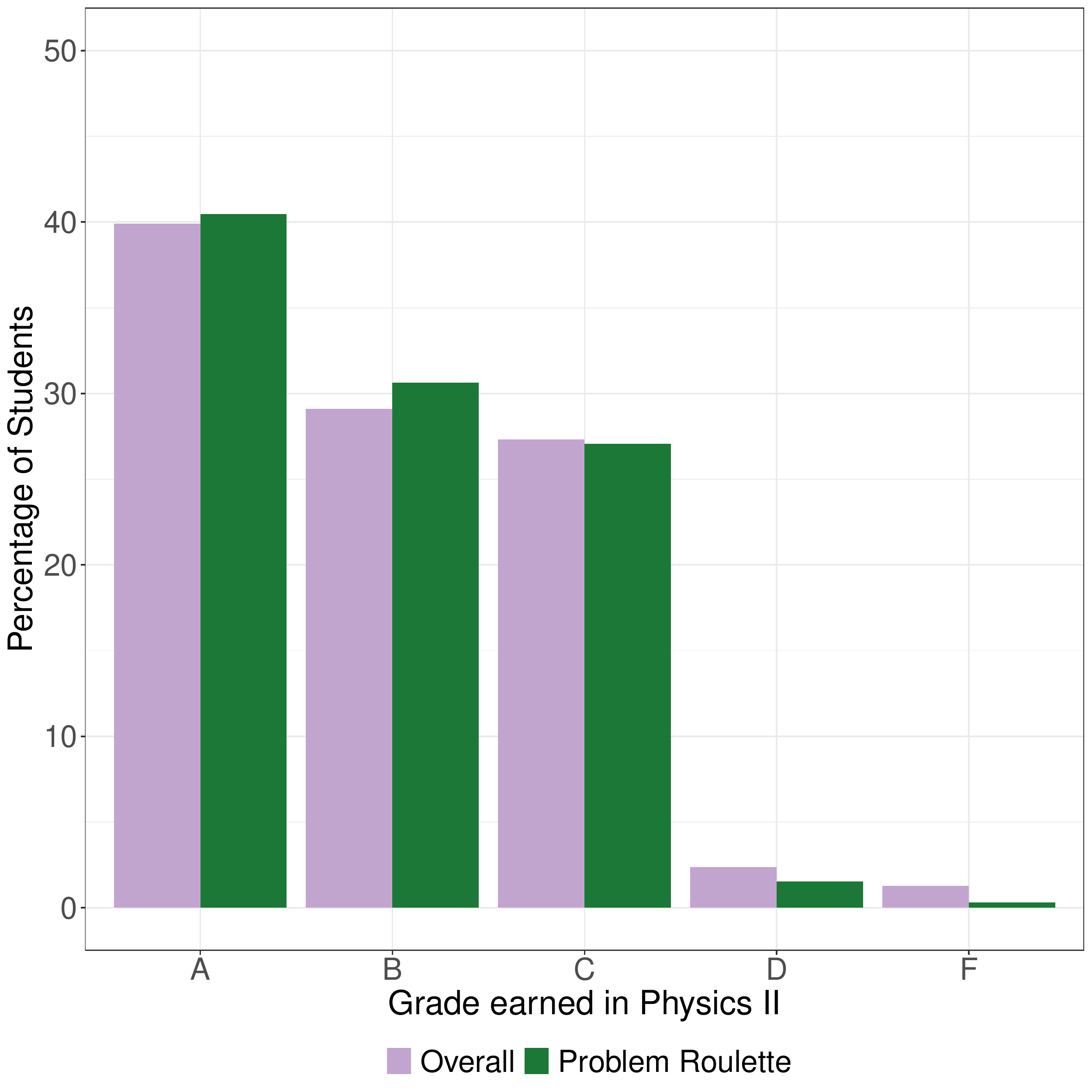}
    \caption{Comparison of the percentage of students who earned each grade in Physics II for students who used our Problem Roulette system and for all students in the course.}
    \label{fig:app_grade}
\end{figure}

Third, we compared the grades students earned in Physics II (Fig. \ref{fig:app_grade}) and find that students who earned ``A"s and ``B"s are slightly overrepresented in the Problem Roulette data compared to everyone who took Physics II at our university.

\begin{figure}
    \centering
    \includegraphics[width=.95\linewidth]{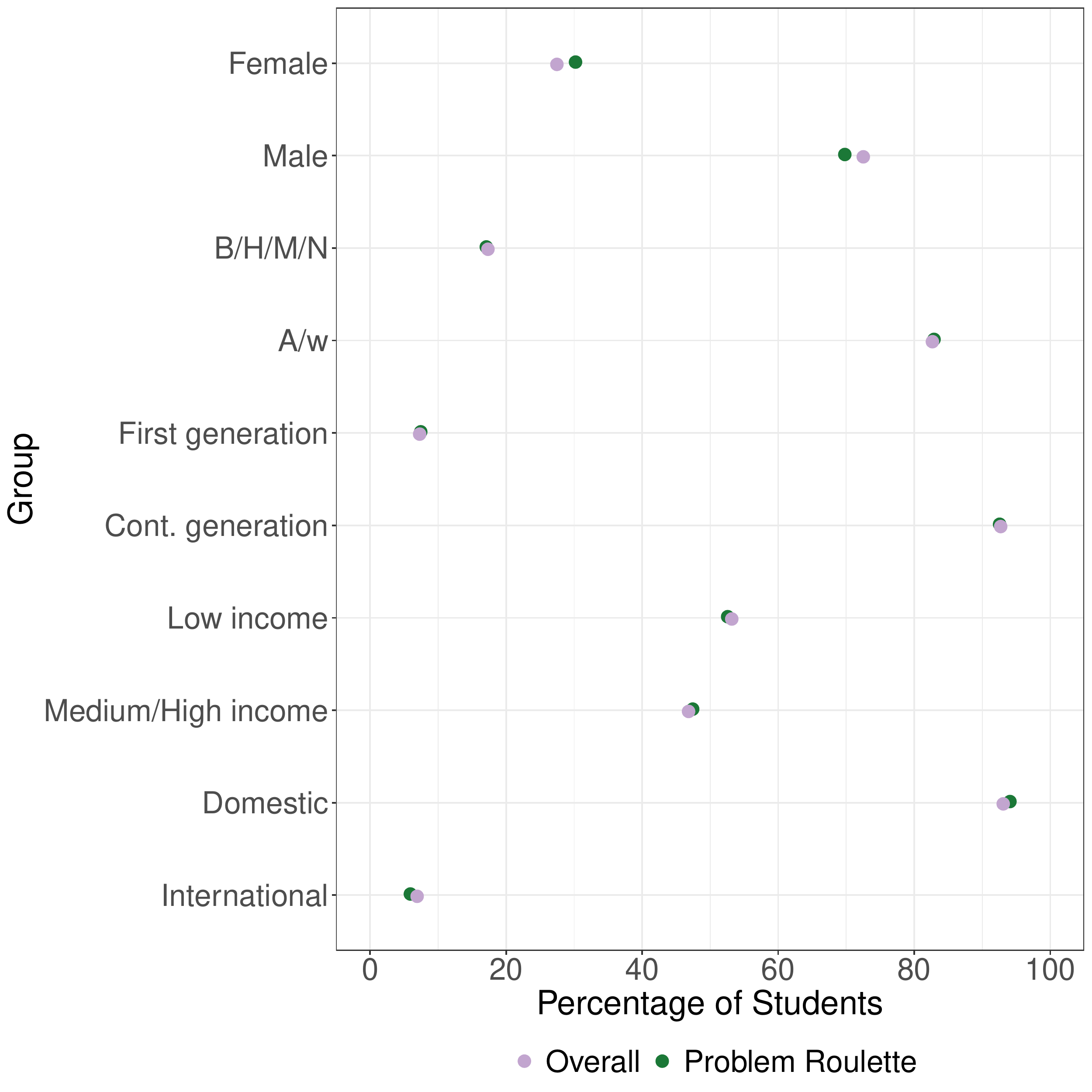}
    \caption{Comparison of the percentage of students based on demographics for students who used our Problem Roulette system and for all students in the course.}
    \label{fig:app_demo}
\end{figure}

Finally, we looked at the demographic distributions of students who used Problem Roulette compared to the overall course (Fig. \ref{fig:app_demo}). We find that for most groups, the students using Problem Roulette are representative of students in the overall course. Sex is the exception to this as female students are slightly overrepresented among Problem Roulette users in this sample.

\bibliography{apssamp}

\end{document}